\documentclass[12pt]{article}
\usepackage{graphicx}
\usepackage{relsize}
\def\babar{\mbox{\slshape B\kern-0.1em{\smaller A}\kern-0.1em
    B\kern-0.1em{\smaller A\kern-0.2em R}}}

\newcommand\pubnumber{ }
\newcommand\pubdate{\today}

\def\napoli{Dipartimento di Fisica, Universit\`a degli Studi di Milano \\
and INFN Sezione di Milano - I-20133 Milano, ITALY}
\def\support{\footnote{On behalf of the \babar\ Collaboration.}}

\def\Title#1{\begin{center} {\Large #1 } \end{center}}
\def\Author#1{\begin{center}{ \sc #1} \end{center}}
\def\Address#1{\begin{center}{ \it #1} \end{center}}

\newcommand\pubblock{\rightline{\begin{tabular}{l} \pubnumber\\
         \pubdate  \end{tabular}}}
\newenvironment{Abstract}{\begin{quotation}  }{\end{quotation}}
\newenvironment{Presented}{\begin{quotation} \begin{center} 
             PRESENTED AT\end{center}\bigskip 
      \begin{center}\begin{large}}{\end{large}\end{center} \end{quotation}}

\def\beq{\begin{equation}}
\def\eeq#1{\label{#1}\end{equation}}
\def\eeqn{\end{equation}}

\def\beqa{\begin{eqnarray}}
\def\eeqa#1{\label{#1}\end{eqnarray}}
\def\eeqan{\end{eqnarray}}

\let\bar=\overbar

\def\Dslash{\not{\hbox{\kern-4pt $D$}}}
\def\dslash{\not{\hbox{\kern-2pt $\del$}}}

\def\msb{{\bar{\ssstyle M \kern -1pt S}}}

\begin{document}
\begin{titlepage}
\pubblock

\vfill

\Title{Measurement of the weak phase $\alpha$ from $B^0 \to a_1(1260)^{\pm} \pi^{\mp}$ decays}
\vfill
\Author{ Simone Stracka\support}
\Address{\napoli}
\vfill
\begin{Abstract}
We present the measurement, performed by the \babar\ Collaboration, of the weak phase $\alpha$ 
from the time dependent $CP$ asymmetries in $B^0 \to a_1(1260)^{\pm} \pi^{\mp}$ decays.
The model error induced by penguin contributions to the $B^0\to a_1(1260)^{\pm}\pi^{\mp}$ channel is estimated from an SU(3) analysis of the branching fractions of $B\to a_1(1260) K$, $B\to K_1(1270)\pi$, and $B\to K_1(1400)\pi$ decays.

\end{Abstract}
\vfill
\begin{Presented}
6th International Workshop on the CKM Unitarity Triangle\\
University of Warwick, United Kingdom, September 6-10, 2010
\end{Presented}
\vfill
\end{titlepage}
\def\thefootnote{\fnsymbol{footnote}}
\setcounter{footnote}{0}

\section{Introduction}

The measurement of the CKM angle $\alpha$ at present-day $B$-factories relies on the
analysis of time-dependent $CP$ violating asymmetries in tree-dominated $b \to u\bar{u}d$ transitions, such as 
$B^0 \to \pi^+\pi^-$, $\rho^+\rho^-$, $\rho^{\pm}\pi^{\mp}$, $a_1(1260)^{\pm}\pi^{\mp}$ (charge-conjugated reactions are implied throughout the text). 
The extraction of $\alpha$ is limited by the penguin contributions 
to the decay amplitude, which shift the value of the phase measured from the time distribution of $B^0$ decays by an amount $\Delta \alpha$ that has to be determined from the experiment. 
One of the strengths of the $B$-factories lies in their ability to use multiple approaches 
to the measurement of $\alpha$, allowing for a better control on model-dependent
estimates of the penguin contributions by comparison with data in many channels. 
Independent measurements of this angle in different channels also help 
to resolve discrete ambiguities that emerge in the extraction of $\alpha$.

The angle $\alpha$ can be measured from $CP$ violating asymmetries in decays of neutral $B$ mesons to non-$CP$ eigenstates~\cite{Aleksan}, such as $\rho^{\pm}\pi^{\mp}$ and $a_1(1260)^{\pm}\pi^{\mp}$. For the $\rho^{\pm}\pi^{\mp}$ final state, $\alpha$ can be extracted without discrete ambiguities by looking at the time-dependent asymmetries in different regions of the $\pi^+\pi^-\pi^0$ Dalitz plot. 
At the present level of statistics, this approach cannot be applied to the four-particle final state resulting from 
$B^0 \to a_1(1260)^{\pm}\pi^{\mp}$ decays. Nevertheless, the analysis of the time-dependent 
$CP$ asymmetries allows to derive an effective value $\alpha_{\rm eff}$, which can be related to $\alpha$ under 
the SU(3) approximate symmetry by measuring the branching fractions of a set of auxiliary $B$ decay channels: $B\to a_1(1260) K$, $B\to K_1(1270)\pi$, and $B\to K_1(1400)\pi$~\cite{Gronau}.
In the following, we report the determination of $\alpha$ 
in the $B^0 \to a_1(1260)^{\pm}\pi^{\mp}$ decays, with the data collected by the \babar\ detector at SLAC.

\section{Branching fraction of $B^0 \to a_1(1260)^{\pm}\pi^{\mp}$ decays}

The $B^0 \to a_1(1260)^{\pm}\pi^{\mp}$ channel was observed  
by \babar\ in 2006~\cite{Palombo06},
by reconstructing the decay of the $a_1(1260)$ axial vector meson (henceforth denoted as $a_1$) into the dominant $\rho\pi$ channel. 
The signal contribution is separated from background by means of an unbinned 
maximum-likelihood (ML) fit to a set of five discriminating variables.
Two kinematic variables, the energy substituted mass $m_{ES}=\sqrt{s/4-p^2_B}$ 
and the energy difference $\Delta E=E_B-\sqrt{s}/2$, where the $B$ 
four-momentum $(E_B,p_B)$ is defined in the $e^+e^-$ center-of-mass (CM) frame, 
allow to discriminate between correctly reconstructed $B$ candidates 
and fake candidates resulting from random 
combination of particles. 
Topological variables, combined into a Fisher discriminant ${\cal F}$, provide 
further distinction between the jet-like shape of 
continuum $e^+e^-\to q\bar{q}$ events ($q=u,d,s,c$), which is the most abundant 
source of background, and the more isotropic $B$ decays. 
The two remaining variables characterize the resonant behavior of the reconstructed 
three-particle system in the final state: the $\pi^{\pm}\pi^-\pi^+$ invariant mass $m_{a_1}$
and the cosine $\cos\theta_H$ of the angle between the momentum of the bachelor pion and 
the normal to the plane described by the resonant three-pion system, in the $a_1$ rest frame. 
The $\cos\theta_H$ distribution allows to discriminate between different $J^P$ hypotheses for the 
three-pion resonance.
The lineshape parameters for the $a_1$ meson are left free in the fit, to minimize systematic uncertainties.

A signal yield of $421\pm 48{\rm (stat.)}$ events is extracted from the fit to the \babar\ data ($218\times 10^{6}$ $B\bar{B}$ pairs), which corresponds to a branching fraction ${\cal B}(B^0\to a_1^{\pm}\pi^{\mp})=(33.2\pm 3.8{\rm (stat.)}\pm 3.0{\rm (syst.)})\times 10^{-6}$, assuming a 50\% branching fraction for
the $a_1(1260)^+\to \pi^+\pi^+\pi^-$ decay~\cite{Palombo06}. 
These results are in good agreement with the branching fraction extracted by Belle, 
$(29.8 \pm 3.2{\rm (stat.)} \pm 4.6{\rm (syst.)})\times 10^{-6}$~\cite{Belle}.

\section{Time-dependence  of $B^0 \to a_1^{\pm}\pi^{\mp}$ decays}

\begin{figure}[htb]
 \centering
\begin{minipage}{0.42\linewidth}
\begin{center}
 \includegraphics[height=2.5in, trim=0.5cm 0 0.5cm 0,clip]{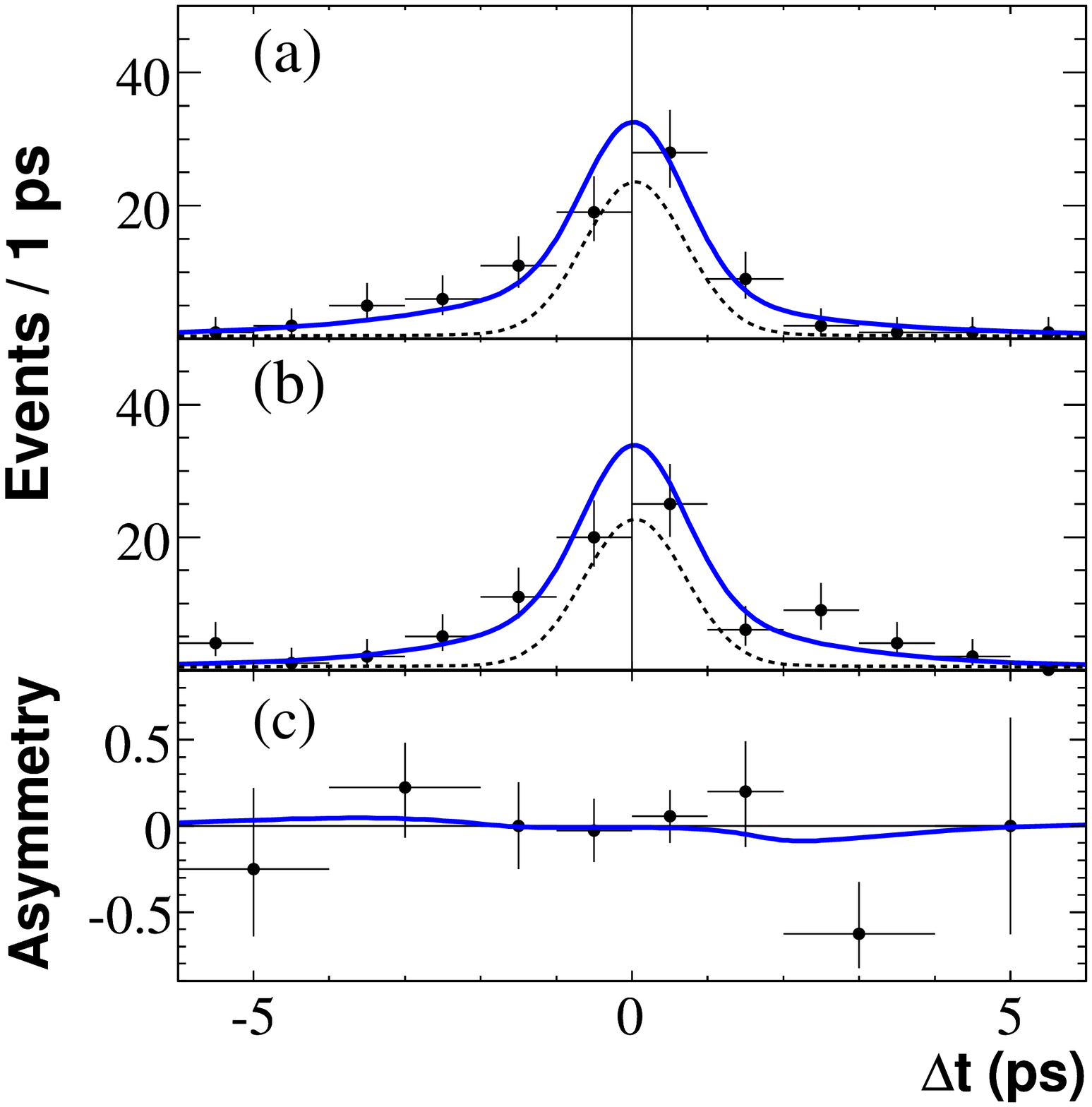}
\end{center}
\end{minipage}
\begin{minipage}{0.30\linewidth}
\begin{center}
 \includegraphics[height=2.9in, trim=13.5cm 0.6cm 0 0cm,clip]{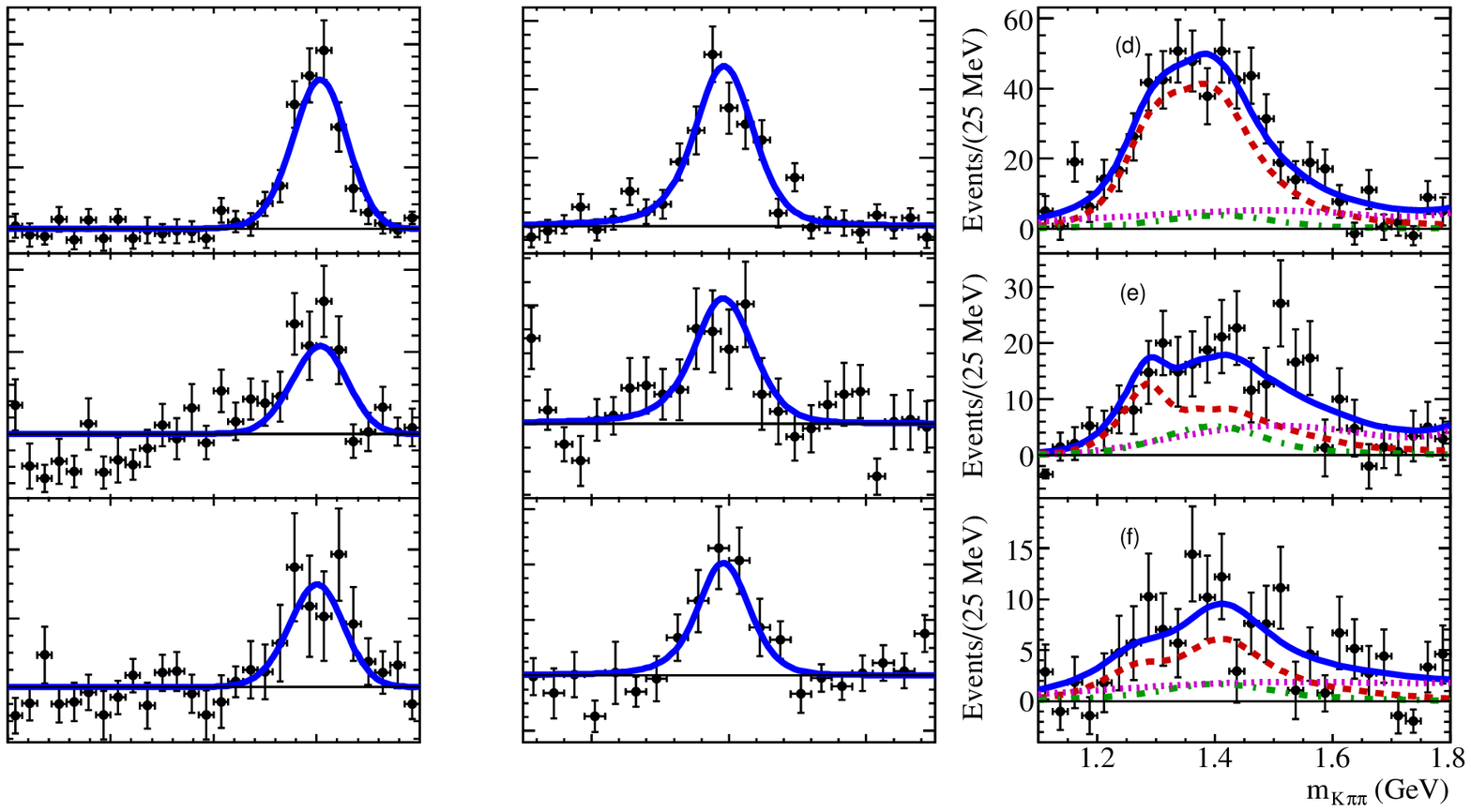}
\end{center}
\end{minipage}
\begin{minipage}{0.26\linewidth}
\begin{center}
 \includegraphics[height=1.4in, trim=0.5cm 0.5cm 0 0.8cm, clip]{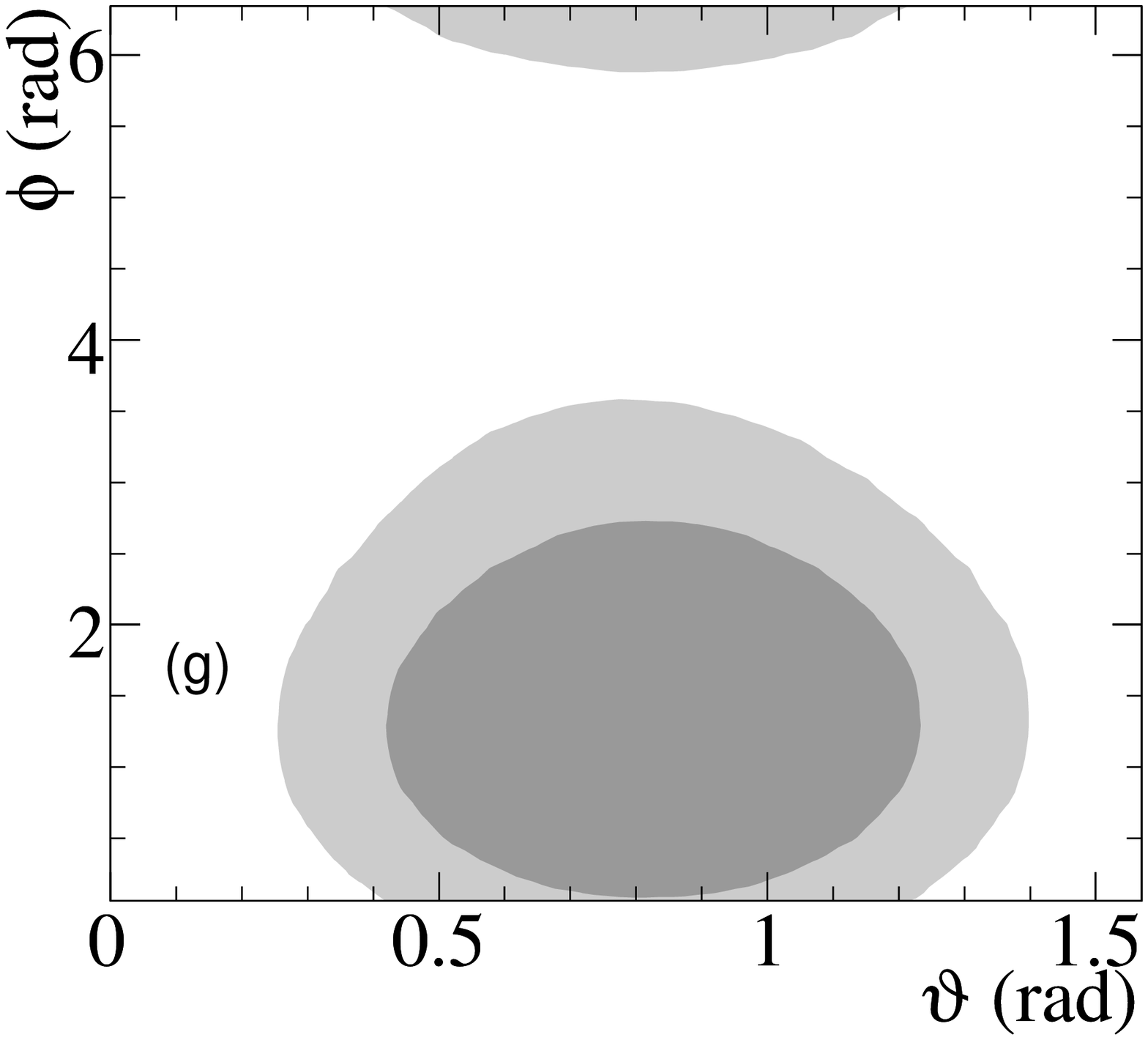}
 \includegraphics[height=1.4in, trim=0.5cm 0.5cm 0 0.8cm, clip]{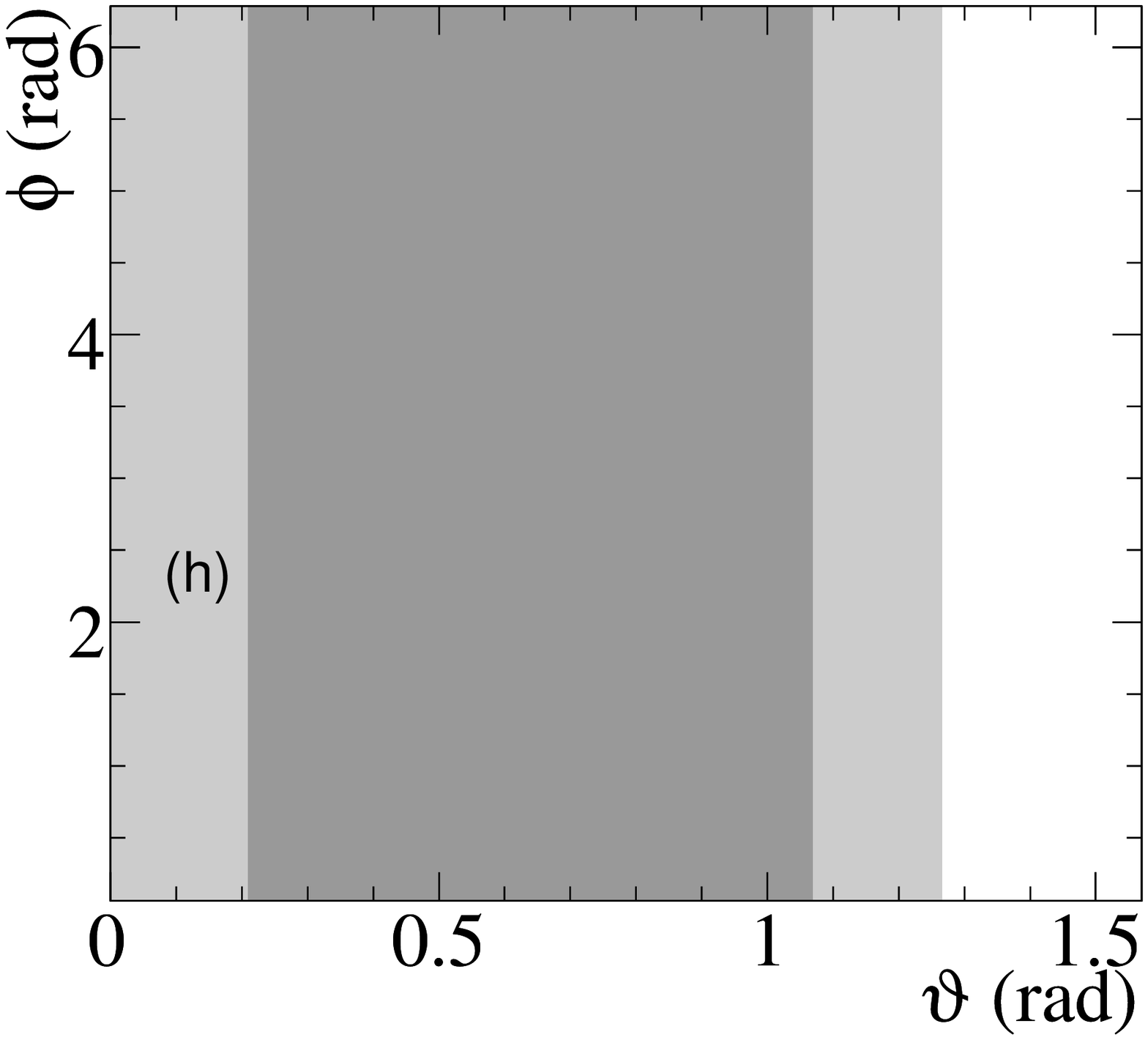}
\end{center}
\end{minipage}
 \caption{\label{fig:a1td} 
(a-c): Projections onto $\Delta t$ of data (points) for (a) $B^0$, (b) $\bar{B}^0$ tags, and (c) the asymmetry between $B^0$ and $\bar{B}^0$ tags~\cite{Lombardo07}. 
(d-f): Continuum-background subtracted projections of the data (points) on  
$m_{K\pi\pi}$ for 
(d,e) $B^0$ and (f) $B^+$ events:  
 (d,f) events with $0.846<m_{K\pi}<0.946\,{\rm GeV}$ and (e) events not included in (d,f) with $0.500<m_{\pi\pi}<0.800\,{\rm GeV}$. The
solid line is the sum of the fit functions for the decay modes $K_1(1270)\pi + K_1(1400)\pi$ (dashed), $K^*(1410)\pi$ (dash-dotted), 
$K^*(892)\pi\pi$ (dotted)~\cite{Stracka09}. 
The dashed curve is normalized to (d) $545$, (e) $245$, and (f) $141$ events.  
(g,h): 68\% (light) and 90\% CL (dark) $\vartheta-\phi$ regions in (g) $B^0$ and (h) $B^+$ decays to $K_1\pi$~\cite{Stracka09}.
}
 \end{figure}

With a sample of $384\times 10^{6}$ $B\bar{B}$ pairs, 
\babar\ performed a ML fit to 29300 selected events, resulting in a signal yield of $608 \pm 53{\rm (stat.)}$ ($461 \pm 46{\rm (stat.)}$ events with their flavor identified), 
and measured the time distribution of the $B^0 \to a_1^{\pm}\pi^{\mp}$ decays 
\begin{equation} \nonumber
f^{a_1^{\pm}}_q(\Delta t) \propto (1\pm {A}_{CP}) \bigg\{1+q \left[ (S\pm\Delta S)\sin(\Delta m_d \Delta t)+ (C\pm\Delta C)\cos(\Delta m_d \Delta t)\right]\bigg\}, 
\end{equation}
where $\Delta m_d = 0.502 \pm 0.007\,\rm{ps}^{-1}$ is the $B^0-\bar{B}^0$ mixing frequency, %
and $q=+1$ ($-1$) if the other $B$ in the event decays as a $B^0$ (${\bar B}^0$).
The observed time-dependent rates and asymmetry are shown in Fig.~\ref{fig:a1td} (a-c), 
and take into account the $\Delta t$ resolution function and the dilution from incorrect flavor assignment. They  correspond to 
${A}_{CP} = -0.07\pm 0.07 \pm 0.02$, $S = 0.37\pm 0.21 \pm 0.07$, $\Delta S = -0.14\pm 0.21 \pm 0.06 $, 
$C = -0.10\pm 0.15 \pm 0.09$, and $\Delta C = 0.26\pm 0.15\pm 0.07$~\cite{Lombardo07}, where the first error is statistical and the second systematic 
(dominated by 
the modeling of the signal distributions and by $CP$ violation in the $B\bar{B}$ background). Linear correlations are at the $O(\%)$ level.
These parameters can be related to the effective value %
$\alpha_{\rm eff}\equiv \alpha + \Delta\alpha$, by the relation
\begin{equation}
 S\pm \Delta S = \sqrt{1+(C\pm\Delta C)^2} \times \sin\left(2\alpha_{\rm eff}\pm \hat{\delta}\right),
\end{equation}
where $\hat{\delta}$ is the strong phase between the tree amplitudes of $B^0$ decays to $a_1^+\pi^-$ and 
$a_1^-\pi^+$. %
The strong phase can be averaged out to yield $\alpha_{\rm eff}$ with an eightfold ambiguity in the range $[0,180]^{\circ}$, which can be reduced by assuming $\hat{\delta}\ll 1$, as suggested by factorization~\cite{Gronau}.
The selected %
solutions are $\alpha_{\rm eff}=(11\pm 7)^{\circ}$ and $\alpha_{\rm eff}=(79\pm 7)^{\circ}$, where the error is 
statistical and systematic combined.

\section{SU(3) analysis and $B \to a_1 K$, $B\to K_1 \pi$}

The effect of penguin pollution $\Delta \alpha = \alpha_{\rm eff} - \alpha$ can be evaluated from 
auxiliary measurements by introducing flavor-symmetry arguments. Since an isospin analysis is not feasible~\cite{Pentagon}, 
our approach is based on a set of SU(3) relations~\cite{Gronau}, that allow to estimate 
the size of penguin amplitudes from the branching fractions of the $\Delta S =1$ partners of the 
$B^0\to a_1^{\pm}\pi^{\mp}$ decays: $B\to a_1 K$ and $B\to K_{1A}\pi$, where the 
$K_{1A}$ state belongs to the same SU(3) octet as the $a_1$ meson. 
This approach is effective because 
in these channels the penguin amplitudes are enhanced by a CKM factor $\bar{\lambda}^{-1}$ 
($\bar{\lambda}\approx 0.23$) with respect to the $\Delta S = 0$ transitions $B\to a_1 \pi$. 
Bounds on $\Delta \alpha$ can be derived from the following ratios of $CP$-averaged rates: 
\begin{eqnarray}\label{ratios}
 R^0_+ \equiv \frac{\bar{\lambda}^2 f^2_{a_1} \bar{{\cal B}}(B^0\to K_{1A}^+\pi^-)}{f^2_{K_{1A}} \bar{{\cal B}}(B^0\to a_1^+\pi^-)}, & \quad  &
R^+_+ \equiv \frac{\bar{\lambda}^2 f^2_{a_1} \bar{{\cal B}}(B^+\to K_{1A}^0\pi^+)}{f^2_{K_{1A}} \bar{{\cal B}}(B^0\to a_1^+\pi^-)},\\
 R^0_- \equiv \frac{\bar{\lambda}^2 f^2_{\pi} \bar{{\cal B}}(B^0\to a_1^- K^+)}{f^2_{K} \bar{{\cal B}}(B^0\to a_1^-\pi^+)}, & \quad  &
R^+_- \equiv \frac{\bar{\lambda}^2 f^2_{\pi} \bar{{\cal B}}(B^0\to a_1^+ K^0)}{f^2_{K} \bar{{\cal B}}(B^0\to a_1^-\pi^+)},
\end{eqnarray}
where the ratios of the decay constants parameterize factorizable SU(3) corrections. Nonfactorizable contributions 
to $\Delta S=0$ and $\Delta S=1$ transitions from exchange and weak annihilation diagrams, respectively, are 
neglected~\cite{Gronau}.
In the above expressions, $f_{a_1}$ and $f_{K_{1A}}$ are obtained from the study of $\tau$ decays~\cite{Decay}. 

 With an analysis similar to the one %
for the $B^0\to a_1^{\pm}\pi^{\mp}$ channel, %
\babar\ measured, from a sample of $383\times 10^{6}$ $B\bar{B}$ pairs, the branching fractions of $B^0\to a_1^- K^+$ and $B^+\to a_1^+ K^0$ decays:  
${\cal B}(B^0\to a_1^- K^+) = (16.4\pm 3.0{\rm (stat.)} \pm 2.4{\rm (syst.)}) \times 10^{-6}$ and 
${\cal B}(B^+\to a_1^+ K^0) = (34.8\pm 5.0{\rm (stat.)} \pm 4.4{\rm (syst.)}) \times 10^{-6}$~\cite{Lombardo08}. 

The $K_{1A}$ is a mixture of the $K_1(1270)$ 
and $K_1(1400)$ axial vector mesons, with a mixing angle $\theta=72^{\circ}$: %
$|K_{1A}\rangle = |K_1(1400)\rangle \cos\theta - |K_1(1270)\rangle \sin\theta$. 
Both resonances decay to $K\pi\pi$ % 
through similar intermediate resonances 
and are characterized by overlapping mass distributions, and %
sizeable interference effects 
are thus %
expected. 
The contribution of the $K_{1A}$ state can be isolated by extracting from data the combined branching fraction of $B$ decays to $K_1(1400)\pi$ and $K_1(1270)\pi$, %
and the relative magnitude ($r\equiv\tan\vartheta$) and phase ($\phi$) of 
$B\to K_1(1270)\pi$ and $B\to K_1(1400)\pi$ amplitudes.

In the analysis recently performed by \babar\ with the final data sample of $454\times 10^{6}$ $B\bar{B}$ pairs~\cite{Stracka09}, the combined $K_1(1270)$ and $K_1(1400)$ signal 
is parameterized in terms of a two-resonance, six-channel $K$-matrix model~\cite{ACCMOR} in the $P$-vector approach~\cite{Aitchison}: the $K$-matrix %
describes the propagation and decay of the $K_1$ resonances, while the $P$-vector %
effectively parameterizes the production of the $K_1$ system, along with a recoiling bachelor pion, in $B$ decays. 
The decay couplings %
and the mass poles %
are determined from the results of the partial wave analysis, %
performed by the ACCMOR Collaboration, 
of the diffractively produced $K\pi\pi$ system~\cite{ACCMOR}. %
The production parameters are extracted from \babar\ data by means of a ML fit to $\Delta E$, $m_{ES}$, ${\cal F}$, 
$\cos \theta_H$, and the invariant mass of the resonant $K\pi\pi$ system ($m_{K\pi\pi}$), which 
provides sensitivity to the individual contributions of the %two 
$K_1$ resonances. 

The continuum-background subtracted $m_{K\pi\pi}$ distribution in data is shown in Fig.~\ref{fig:a1td} (d-f). 
Including systematic uncertainties, dominated by the effect of interference between the $K_1$ and the non-resonant components, the combined signal branching fractions are ${\cal B}(B^0\to K_1(1270)^+\pi^- + K_1(1400)^+\pi^-)=31^{+8}_{-7}\times 10^{-6}$ and 
${\cal B}(B^+\to K_1(1270)^0\pi^+ + K_1(1400)^0\pi^+)=29^{+29}_{-17}\times 10^{-6}$. %
The information about the fraction and phase of the two resonances (Fig.~\ref{fig:a1td} (g,h)) is used %
to calculate ${\cal B}(B^0\to K_{1A}^+\pi^-)=14^{+9}_{-10}\times 10^{-6}$ and ${\cal B}(B^+\to K_{1A}^0\pi^+)<36\times 10^{-6}$, where the latter upper limit is evaluated at the 90\% confidence level (CL)~\cite{Stracka09}. 

\section{Bounds on $\Delta \alpha$}

The bounds on $|\Delta\alpha|$ are derived by inverting the relations~\cite{Gronau}
\begin{eqnarray}
\cos 2(\alpha_{\rm eff} \pm \hat{\delta} - \alpha) & \ge & (1-2R^0_{\pm})/\sqrt{1-(A_{CP}^{\pm})^2}, \\
\cos 2(\alpha_{\rm eff} \pm \hat{\delta} - \alpha) & \ge & (1-2R^+_{\pm})/\sqrt{1-(A_{CP}^{\pm})^2}.
\end{eqnarray}
A Monte Carlo method is used to derive the 68\% and 90\% CL upper limits for the bounds: replicas  
of the input quantities are generated from the experimental distributions, and for each simulated set 
of values the above system of inequalities is solved. % 
This study yields the bound $|\Delta\alpha|<11^{\circ}~(13^{\circ})$ at the 68\% (90\%) CL, and 
the final result $\alpha = (79 \pm 7 \pm 11)^{\circ}$ for the solution compatible with the CKM global fits, 
where the first error is statistical and systematic combined and the second is %a model error 
due to penguin pollution.  

The presence of non-factorizable SU(3) breaking effects can be tested, e.g., at LHCb or at a Super $B$-factory, 
by studying auxiliary decay channels such as $K_1^{\pm}K^{\mp}$. The impact of such corrections can be estimated by writing $p'_{\pm}= -c_{\pm}\bar{\lambda} p_{\pm}$, where $p'_{\pm}$ ($p_{\pm}$) is the penguin amplitude in $\Delta S=1$ ($\Delta S=0$) transitions, and the departure of $c_{\pm}$ from $1$ %
quantifies the amount of non-factorizable SU(3) breaking. % 
For $c_{\pm}=0.7$, the bounds are expected to increase by about $3^{\circ}$. 
Finally, a full SU(3) fit may provide an experimental test of $\hat{\delta}\ll 1$.

\section{Conclusions}
\babar\ has measured the CKM angle $\alpha$ from the $B^0\to a_1^{\pm}\pi^{\mp}$ channel, 
and has obtained a value $\alpha = (79 \pm 7 \pm 11)^{\circ}$. 
This independent determination of $\alpha$ is 
consistent with the world average of the $\rho\rho$, $\rho\pi$, and $\pi\pi$ channels and with CKM global fits.

\end{document}